\documentclass[numberedappendix]{emulateapj}

\usepackage{graphicx}% Include figure files
\usepackage{bm}% bold math
\usepackage{epsf}
\usepackage{verbatim}
\usepackage{amsmath}
\usepackage{amsfonts}
\usepackage{epsfig}
\usepackage{dcolumn}
\usepackage{graphicx}
\usepackage{amssymb}

\usepackage{xcolor}
\definecolor{darkgreen}{rgb}{0.0,0.5,0.0}

\usepackage[colorlinks,linkcolor=darkgreen,citecolor=darkgreen]{hyperref} % Works with PDFLaTeX

\def\MyMNRAS#1{}

\newcommand{\eg}{\emph{e.g.,} }

\newcommand{\be}{\begin{equation}}
\newcommand{\ee}{\end{equation}}
\newcommand{\bea}{\begin{equation*}}
\newcommand{\eea}{\end{equation*}}
\newcommand{\beqr}{\begin{eqnarray} \nonumber}
\newcommand{\eeqr}{\end{eqnarray}}
\newcommand{\beqrb}{\begin{eqnarray}}
\newcommand{\eeqrb}{\nonumber \end{eqnarray}}
\newcommand{\fin}{\mbox{ .}}
\newcommand{\coma}{\mbox{ ,}}

\newcommand{\cm}{\mbox{ cm}}
\newcommand{\sr}{\mbox{ sr}}
\newcommand{\se}{\mbox{ s}}

\newcommand{\erg}{\mbox{ erg}}

\newcommand{\kpc}{\mbox{ kpc}}

\newcommand{\keV}{\mbox{ keV}}

\newcommand{\GeV}{\mbox{ GeV}}
\newcommand{\TeV}{\mbox{ TeV}}
\newcommand{\PeV}{\mbox{ PeV}}

\newcommand{\muG}{\mbox{ $\mu$G}}

\newcommand{\gama}{$\gamma$}

\newcommand{\myNi}{\emph{(i)}\,}
\newcommand{\myNii}{\emph{(ii)}\,}
\newcommand{\myNiii}{\emph{(iii)}\,}

\newcommand{\dgr}{{^\circ}}
\newcommand{\till}{{\mbox{--}}}

\defcitealias{SuEtAl10}{S10}
\defcitealias{FermiBubbles14}{F14}
\defcitealias{IceCube37events}{A14}
\defcitealias{IceCube53EventsShort}{K15}
\defcitealias{KeshetGurwich17a}{KG17a}
\defcitealias{KeshetGurwich17b}{KG17b}
\defcitealias{newlimit}{F17}

\newcommand{\Su}{{\citetalias{SuEtAl10}}}
\newcommand{\FT}{{\citetalias{FermiBubbles14}}}
\newcommand{\Aa}{{\citetalias{IceCube37events}}}
\newcommand{\Ko}{{\citetalias{IceCube53EventsShort}}}
\newcommand{\KGa}{{\citetalias{KeshetGurwich17a}}}
\newcommand{\KGb}{{\citetalias{KeshetGurwich17b}}}
\newcommand{\Fa}{{\citetalias{newlimit}}}

\newcommand{\ApJMark}[1]{{#1}}

\begin{document}

\title{IceCube constraints on the Fermi Bubbles}

\shorttitle{Fermi bubble neutrinos}

\author{Nimrod Sherf\altaffilmark{1}, Uri Keshet\altaffilmark{1}, and Ilya Gurwich\altaffilmark{2}}

\altaffiltext{1}{Physics Department, Ben-Gurion University of the Negev, POB 653, Be'er-Sheva 84105, Israel; sherfnim@post.bgu.ac.il, ukeshet@bgu.ac.il}

\altaffiltext{2}{Department of Physics, NRCN, POB 9001, Beer-Sheva 84190, Israel, gurwichphys@gmail.com}

\shortauthors{Sherf et al.}

\date{\today}

\begin{abstract}
We analyze the IceCube four-year neutrino data in search of a signal from the Fermi bubbles.
No signal is found from the bubbles or from their dense shell, even when taking into account the softer background.
This imposes a conservative $\xi_i<8\%$ upper limit on the cosmic-ray ion (CRI) acceleration efficiency, and an $\eta\equiv \xi_e/\xi_i \gtrsim0.006$ lower limit on the electron-to-ion ratio of acceleration efficiencies (at the $2\sigma$ confidence level). For typical $\xi_i$, a signal should surface once the number of IceCube neutrinos increases by $\sim$an order of magnitude, unless there is a $<$PeV cutoff on the CRI spectrum.
\end{abstract}

\keywords{neutrinos, gamma rays: ISM, (ISM:) cosmic rays, Galaxy: center, acceleration of particles}

\maketitle

\section{Introduction}
\label{sec:Intro}

A pair of highly extended, double-lobed $\gamma$-ray bubbles was identified \citep[][henceforth \Su]{DoblerEtAl10, SuEtAl10} using the Fermi-LAT data.
These so-called Fermi bubbles (FBs) extend $\sim 50^\circ$ above and below the Galactic plane and their location and morphology suggest an association with the Galactic center.
They show an approximately flat (constant $\epsilon_\gamma^2 dN_\gamma/d\epsilon_\gamma$) spectrum across the energy range of 1--100 GeV.
Their flux, $\sim 5\times 10^{-7}\GeV \se^{-1} \cm^{-2} \sr^{-1}$ at this energy range, is nearly uniform spatially, and corresponds to a luminosity of $L_\gamma\simeq 4\times 10^{37}\erg\se^{-1}$ \citep[assuming a $d\simeq 10\kpc$ distance; {\Su} and][henceforth \FT]{FermiBubbles14}.

The FBs show counterparts in other bands: in X-rays, a $\lesssim \keV$ shell \citep[][henceforth \KGb]{BlandHawthornAndCohen03, KeshetGurwich17b}; in microwaves, the so-called microwave haze \citep{Finkbeiner04}; and possibly in the radio, an extended polarization signal \citep{CarrettiEtAl13}. The combined electromagnetic signature, along with additional evidence \citep{FoxEtAl15, MillerBregman16}, indicate a supersonic outflow from an explosive event near the Galactic center several megayears ago. The FB edges then trace a strong, outgoing, forward shock, accelerating high-energy cosmic-ray (CR) ions (CRIs) and electrons (CREs), which subsequently diffuse into the FBs \citep[][henceforth \KGa]{KeshetGurwich17a}.
Other models include a  starburst \citep{CarrettiEtAl13, Lacki14, SarkarEtAl15}, central massive black hole jets \citep{ChengEtAl11, GuoMathews12, GuoEtAl12, MouEtal14} or outflows \citep{ZubovasNayakshin12}, and steady star-formation \citep{CrockerEtAl15}.

The \gama-ray emission from the FBs has been modeled as either leptonic \citep[\Su;][and \FT]{YangEtAl13} or hadronic \citep[\Su;][and \FT]{CrockerAharonian11, FujitaEtAl13}.
Neither of these models were able to simultaneously account for both \gama-ray and microwave signals without invoking ad hoc, physically unmotivated cutoffs on the CR spectrum.
In a separate publication (I. Gurwich \& U. Keshet 2017, in preparation), we argue that the only natural FB model fitting the data is leptonic, and present a self-consistent, natural, single-zone model with no ad hoc cutoffs; this leads to a $\sim 10\GeV$ cooling break in the CRE spectrum, yielding a $\sim 1 \GeV$ break in the \gama-rays. The CRI spectrum, on the other hand, is not constrained here by the \gama-rays.

In the absence of an unnatural cutoff on the spectrum of the CRIs, their acceleration in the FBs should proceed to high energies of the order of \citep{Hillas}
\begin{eqnarray}
E_{max} & \simeq & Ze \beta B L \\
& \simeq & 500 \left(\frac{ZB}{5\muG}\right)\left(\frac{\beta}{0.01}\right)\left(\frac{L}{10\kpc}\right) \PeV \coma \nonumber
\end{eqnarray}
where $B$, $L$, and $\beta$ are the typical FB magnetic field, length scale, and shock velocity normalized to the speed of light, respectively, and $Ze$ is the CRI charge.
Inelastic collisions of these CRIs with the ambient gas should therefore generate neutrinos of energies up to at least $\sim 20\PeV$, well within the IceCube band \citep{IceCubeManual,IceCubeIntro, AhlersMurase14}.

Here we examine the neutrino data presently available from IceCube, in search of counterpart FB neutrinos (FB$\nu$s).
The manuscript is arranged as follows.
In \S\ref{sec:data} we introduce the IceCube data used in our study.
The analysis is presented in \S\ref{sec:analysis}, and its implications for the CR acceleration are outlined in \S\ref{sec:Implications}. The results are summarized and discussed in \S\ref{sec:discussion}.

\section{Data preparation}
\label{sec:data}

We use the IceCube four-year data set, containing 54 high-energy, $\sim 10\TeV\till 2\PeV$ neutrino events
\citep[][henceforth {\Aa} and \Ko]{IceCube37events, IceCube53EventsShort}.
The data can be broadly classified into two types of events: showers (or cascades), which are mainly attributed to electron and tau neutrinos; and tracks, which are primarily associated with muon neutrinos \citep{IceCube28events}.

The effective area $A(\bm{\omega})$ of IceCube depends on the direction $\bm{\omega}$ of the impacting neutrinos, in particular their declination. This dependence is due to the event selection process and the natural asymmetry of the detector caused by the Earth \citep{IceCube28events}.
Here and below, the boldface symbols are unit vectors.
The effective area also depends on energy, but here we use the estimated effective area after integration over energies above $60\TeV$.

\begin{figure*}
\centerline{\epsfxsize=18cm
%\epsfbox{IceCubeFig6.eps} %highres
\epsfbox{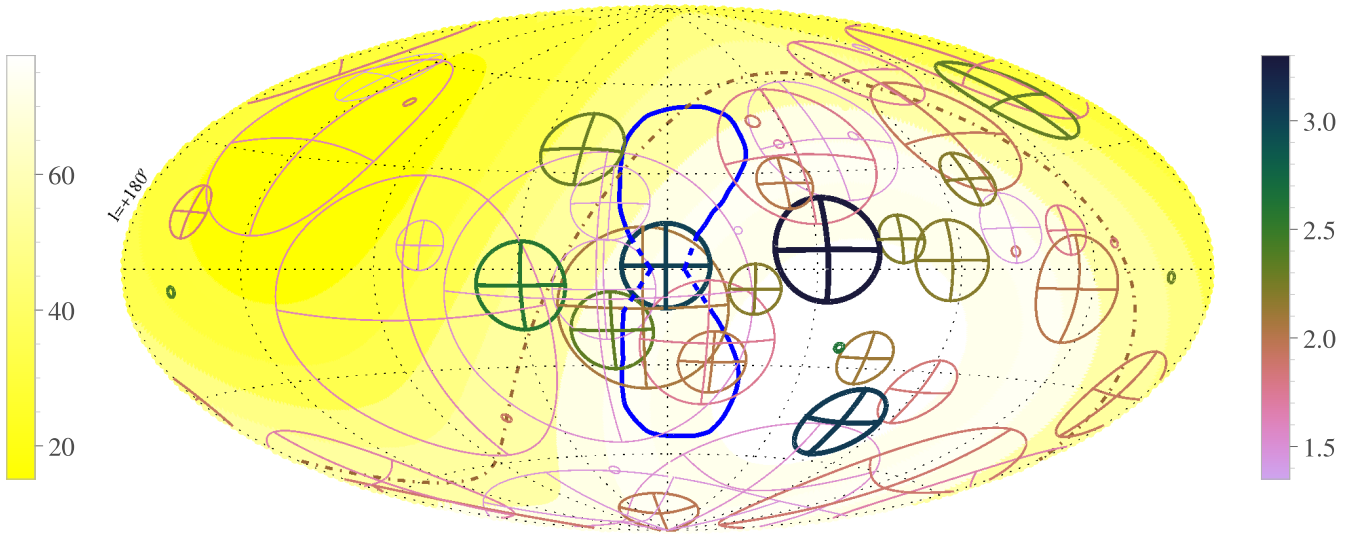} %IceCubeFig7R.eps} %lowres
}
	\caption{\label{fig:Map}
    IceCube neutrino localizations ($1\sigma$ circles, with crosses for the shower events) and energies (line widths and right color bar give $\log_{10}(\epsilon/\mbox{TeV})$), shown in Galactic coordinates with a Hammer--Aitoff projection, superimposed on the effective area of IceCube for a flat spectrum (left color bar in m$^2$ units).
    The FBs, shown using the edges of {\KGa} (edges 1 therein; solid blue contours; extended to the Galactic plane $(l=\pm5\dgr,b=0)$ as shown by the dashed lines), lie mostly south of the celestial equator (dot-dashed brown curve).
   	}
\end{figure*}

The events are based on analyses with an isotropic event selection and with a containment cut meant to remove atmospheric events.
Neutrino scattering in the Earth leads to a bias toward the southern hemisphere, where the total event rate is nearly isotropic
(N. Kurahashi Neilson and N. Whitehorn 2016, private communications), to within $\sim10\%$ accuracy (based on the simulated event rate as a function of declination for an isotropic, spectrally-flat astrophysical neutrino background; see \Ko, Figure 4).
The inferred effective area is shown (in shades of yellow) in Figure \ref{fig:Map}.

Out of the 54 neutrino events, including both showers and tracks, events 1--37 are taken from \Aa, and events 38--54 are from  \Ko.
Two out of these 54 events (events number 28 and 32) were excluded from the analysis because they probably arise from CR air showers (\Aa).
Figure \ref{fig:Map} shows the remaining 52 neutrinos in Galactic coordinates, using a Hammer--Aitoff projection.

We model the arrival direction $\bm{\omega}$ of each event $j$ with a Fisher distribution,
\begin{equation}
f_j(\bm{\omega})
=\frac{1}{4\pi\sigma_j^2\sinh(\sigma_j^{-2})}\exp\left(\frac{\bm{\omega}\cdot\bm{n}_j}{\sigma_j^2}\right) \coma
\end{equation}
centered upon the tabulated most likely coordinates $\bm{n}_j$ in {\Aa} (Supplementary Table 1) and in {\Ko} (Table 1), with dispersion $\sigma_j$ given by the event's tabulated angular error; the effects of the variable effective area on the probability distribution of the arrival direction are neglected.
The tabulated neutrino energies (line thickness and color in Figure \ref{fig:Map}) are used to estimate the flux.

\section{Analysis}
\label{sec:analysis}

The FB edges have been previously traced using gradient filters of different scales ({\KGa}; see Figure \ref{fig:Map}), as well as manually (\eg \Su).
Different edge tracing methods differ by several degrees, and up to $15\dgr$ in the western longitudes, far from the Galactic plane.
Here we use edge 1 of {\KGa}, which is based on a gradient filter on $6\dgr$ scales.
This is sufficient for tracing the FB edge at latitudes $|b|\gtrsim10\dgr$.
In order to extend the edges deeper into the Galactic plane, we arbitrarily extend the east (west) edge to a longitude of $+5\dgr$ ($-5\dgr$) at the plane.
This slightly overestimates the size of the FBs near the Galactic center, so we may only impose an upper limit on the FB$\nu$s.

The number of neutrinos estimated to arrive from some region of interest $\Omega_0$ in the sky is given by
\begin{equation} \label{eq:NOmega0}
N(\Omega_0) = \int_{\Omega_0} d\bm{\omega} \sum_j f_j(\bm{\omega}) \fin
\end{equation}
We wish to determine if, within $\Omega_0$ (namely, the full-sky or the southern equatorial hemisphere), there are excess neutrinos arriving from a target $\Omega$ (namely, the FBs or some part of them).
We thus compare the estimated number of neutrinos arriving from the overlap between the region of interest and the target, $N(\Omega\cap\Omega_0)$, with the null-hypothesis of isotropically distributed neutrinos within $\Omega_0$ with no additional signal from $\Omega$,
\begin{equation} \label{eq:NIso}
N_{iso}(\Omega|\Omega_0) = N(\Omega_0) \frac{\int_{\Omega\cap\Omega_0} A(\bm{\omega}) d\bm{\omega} }{\int_{\Omega_0} A(\bm{\omega}) d\bm{\omega} }\fin
\end{equation}
A signal would correspond to a large number $N(\Omega\cap\Omega_0)$ of target neutrinos, significantly exceeding the no-signal expectation $N_{iso}(\Omega|\Omega_0)$.
Table \ref{tab:summary} summarizes the results of analyzing different combinations of FB targets $\Omega$ inside regions of interest $\Omega_0$.

We start with an all-sky, $\Omega_0=4\pi$ analysis.
For an isotropic neutrino background, out of the 52 IceCube neutrinos, on average $N_{iso}(\Omega|\Omega_0)\simeq 4.6$ would coincidentally arrive from the direction of the FBs, when taking into account the variable effective area, according to Eqs.~(\ref{eq:NOmega0}) and (\ref{eq:NIso}).
The number of IceCube neutrinos actually estimated as arriving from the FB direction is found to be smaller: $N(\Omega\cap\Omega_0)\simeq 3.7$.
We conclude that no signal from the FBs as a whole is seen in a full-sky analysis.

\begin{table*}
\begin{center}
\caption{\label{tab:summary} Tests for FB$\nu$ signals
}
\begin{tabular}{ccccccc}
\hline
$\Omega_0$ & $\Omega$ & $N(\Omega_0)$ & $N_{iso}(\Omega|\Omega_0)$ & $N(\Omega\cap\Omega_0)$ & $E_{iso}(\Omega|\Omega_0)$  & $E(\Omega\cap\Omega_0)$ \\
(1) & (2) & (3) & (4) & (5) & (6) & (7) \\
\hline
All-sky & FBs & 52 & 4.6 %4.4
& 3.7 & 794
%772
& 915 \\
South & FBs & 34.9 & 4.5 %4.5
& 3.6 & 930
%933
& 903 \\
All-sky & shells & 52 & 1.7 %1.7
& 1.3 & 294
% 289
& 269 \\
South & shells & 34.9 & 1.6 %1.6
& 1.2 & 327
%327
& 258 \\
All-sky\tablenotemark{\textdagger} & FBs\tablenotemark{\textdagger} & 51 & 4.5
% 4.4
& 3.2 & 702
% 683
& 343 \\
\hline
\end{tabular}
\tablecomments{
Columns: (1) region of interest; (2) target; (3) number of events in region of interest; (4) isotropic expectation number of events in target; (5) number of events in target; (6) isotropic expectation energy, in TeV; and (7) energy in target, in TeV.\\
{\bf \textdagger} Without event 14, lying near the Galactic center.
}
\end{center}
\end{table*}

The FBs are located almost entirely in the southern equatorial hemisphere, as shown in Figure \ref{fig:Map}.
The larger effective area in this hemisphere suggests less foreground contamination, so we consider it as a second, $\Omega_0=2\pi$ region of interest.
Here, south of the celestial equator, the numbers of neutrinos received from the direction of the FBs is still smaller than that of the no-signal expectation value (see Table \ref{tab:summary}).
Thus, we find no evidence for uniform neutrino emission from the FBs, not even when limiting the region of interest to the southern equatorial hemisphere.

The gas density is thought to be elevated in a thin shell behind the FB edges, so next we consider possible neutrino emission from this shell.
We define the shell as the region inside the FBs at a distance of no more than $5\dgr$ from the edge, and at least $5\dgr$ from the Galactic plane.
Repeating the above analysis for the shell gives, as expected, a smaller value of the isotropic expectation $N_{iso}(\Omega|\Omega_0)$.
However, the number $N(\Omega)$ of neutrinos estimated from the direction of the shell is even smaller.
These results, found both all-sky and in the southern equatorial hemisphere, do not indicate a deviation from an isotropic distribution.
Therefore, we cannot identify any association between the IceCube neutrinos and the part of the sky subtended by the FBs, neither when assuming uniform emission from the bubbles, nor when examining enhanced emission from their shell.

As the spectrum of the IceCube neutrinos \citep{IceCube15_SoftBackground} is softer than what is expected from the FBs, we consider weighing the neutrino events according to their energies. For example, one may search for a signature of the FBs in the total deposited neutrino energy rather than in the number of events.
Repeating the above analysis for the deposited energy yields similar results (see Table \ref{tab:summary}).
For an all-sky analysis, the energy of the neutrinos arriving from the direction of the FBs exceeds the isotropic expectation by $\sim20\%$, while that from the direction of the shell is $\sim10\%$ lower. The main cause for the elevated energy from the full FBs can be traced to a single very energetic event (number 14 in \Aa; Supplementary Table 1) near the Galactic center.
Indeed, removing this event from our analysis entirely eliminates the excess energy.
In summary, even using the deposited energy, we are still unable to associate IceCube neutrinos with the FBs.

\section{Implications of No FB$\nu$ Detection}
\label{sec:Implications}

The preceding discussion shows no significant association between IceCube neutrinos and the FBs.
This imposes:
\myNi an upper limit on the FB$\nu$ flux, $F_{FB\nu}$;
\myNii an upper limit on the efficiency $\xi_i$ of the CRI acceleration;
and, using an inverse-Compton (IC) model for the \gama-ray flux,
\myNiii a lower limit on the electron-to-ion ratio, $\eta$, of the CR acceleration efficiencies.

We consider the neutrino energy range of $60 \TeV\till 3 \PeV$.
The logarithmic IceCube flux in this range is $\epsilon^2 dF_{\nu}/d\epsilon \simeq (8 \pm 3) \times 10^{-9} \GeV \se^{-1} \cm^{-2} \sr^{-1}$, computed per neutrino flavor assuming a spectral index of $s=2$ (henceforth); a similar result is obtained for $s=2.5$. Assuming an isotropic astrophysical background, we expect $\sim 4.5$ neutrinos to arrive from the FB region. Poisson statistics with the aforementioned flux then dictates the $2\sigma$ limit,
\begin{equation} \label{eq:upperlimit}
\epsilon^2 \frac{dF_{FB\nu}}{d\epsilon} < 2\times 10^{-8} \GeV \se^{-1} \cm^{-2} \sr^{-1}
\coma
\end{equation}
inclusive for all flavors.

Notice that the logarithmic  $\gamma$-ray flux from pp collisions is proportional to the neutrino flux \citep[\eg][]{rationumbers}, such that one may impose the corresponding limit of
\begin{equation} \label{gammaflux}
\epsilon_\gamma^2 \frac{dF_{\gamma}}{d\epsilon_\gamma} \simeq
\frac{2}{3} \epsilon^2 \frac{dF_{\nu}}{d\epsilon}
\lesssim 10^{-8}  \GeV \se^{-1} \cm^{-2} \sr^{-1}
\end{equation}
at similar photon energies.
Assuming that this constraint extends down to LAT energies, it validates a leptonic origin for the FB \gama-rays.

Taking the inclusive cross section for $\pi^\pm$ production at $\sim\PeV$ energies as \citep{DermerCrossSection, BlattnigEtAl00}
$\sigma=0.3\sigma_{0.3}\mbox{ barn}$, and the fraction of the CRI energy transferred into neutrinos as $0.15$, the upper limit on CRIs becomes
\begin{eqnarray} \label{crpsenergy}
U_{CRI} < 4 \times 10^{54} \sigma_{0.3}^{-1}n_{-3}^{-1}   \erg \fin
\end{eqnarray}
where here we approximate each FB as a semimajor axis $a=5\kpc$, semiminor axis $b=3\kpc$ prolate spheroid, 
centered at a height $5\kpc$ above the Galactic center (which is assumed to lie at a distance $8.5\kpc$ from us), 
with a volume-averaged hydrogen number density of $n=10^{-3}n_{-3}\cm^{-3}$ (\KGb).
We also approximate (henceforth) all of the CRIs as protons.

Assuming an ion temperature, mass-averaged over the FB, of $k_B T\simeq T_{keV} \keV$, the CRI acceleration efficiency $\xi_i$, defined as the fraction of downstream thermal energy deposited in the CRIs, is bounded by
\begin{equation}
\xi_i < 0.08 n_{-3}^{-2} T_{kev}^{-1} \sigma_{0.3}^{-1} \fin
\end{equation}
Taking into account evidence \citep{FoxEtAl15} that the FB ions are significantly hotter than $1 \keV$ (\KGb), indicates a very low ion acceleration efficiency.
It should be noted that the CRE acceleration efficiency $\xi_e$ (defined similarly, as the fraction of downstream thermal energy deposited in the CREs) is also known to be very low in the FBs.

\ApJMark{The CRE spectrum can be evaluated by interpreting the FB \gama-ray spectrum (\FT) below a GeV as arising from IC scattering off the CMB. We then constrain the energy ratio between CREs and CRIs to be}
\begin{equation}
\eta \equiv   \frac{dN_e/dE}{dN_i/dE} \approx \frac{\xi_e}{\xi_i} >0.006 ~\sigma_{0.3}n_{-3}\fin
\end{equation}
\ApJMark{This represents the ratio of the energies deposited in the accelerated  CREs and CRIs. Note that due to the cooling of high-energy CREs, it is not equivalent to the ratio between the present energies contained in these CR species.}

\section{Summary and Discussion}
\label{sec:discussion}

We analyze the four-year IceCube data in search for a signal from the FBs.
We search for a signal both from the FBs as a whole, and as limb-brightened emission from a thin shell behind the FB edges.
We also account for the harder FB spectrum by searching for an energy-weighted signal.

With the presently available data, none of the above tests indicate a significant detection of FB$\nu$s, as summarized in Table \ref{tab:summary}.
This imposes constraints on the FB$\nu$ flux, and therefore on the CRI density above $\sim \PeV$.
Unless the FB CRI spectrum at lower energies features a break or a cutoff, our analysis yields significant constraints on the FB CRI acceleration efficiency $\xi_i$ and on the ratio $\eta$ between CR electron and ion acceleration efficiencies.

The resulting upper limit on $\xi_i$ is $\sim 10\%$ for an average FB temperature of $1\keV$. However, higher temperatures, inferred (\KGb) from the high velocity \citep{FoxEtAl15} of the gas inside the FBs, suggest a low, $<2\%$ efficiency.
Assuming that $\xi_i$ in the FBs is not significantly lower than what is typically estimated, this result implies that either \myNi further observations will detect the FB$\nu$ signal, once the number of detected neutrino events increases by $\sim$ an order of magnitude; or \myNii there is a break or cutoff in the FB CRI spectrum below $\sim\PeV$.

The CR electron-to-ion ratio at a given energy is constrained as $\eta\gtrsim0.006$, consistent with some of the previous models and estimates, but not with others, in particular hadronic emission models  \citep[see, for example, ][]{SNRratio2010, MorlinoCaprioli12}. The validity of this result too depends on the lack of a spectral break or cutoff below a $\PeV$.

While finalizing this project, we were made aware of a similar work \citep[][henceforth {\Fa}]{newlimit}, imposing constraints on CRs from the FBs using the IceCube and HAWC data.
Their results are quite similar to ours overall.
The FB$\nu$ flux upper limit of {\Fa} is $\sim 50\%$ higher
than ours, whereas the upper limits on the CRI agree even better, to within 20\% (when corrected for the different gas densities used in the two papers), despite {\Fa} relying on HAWC, rather than IceCube, to obtain this constraint.

The limits on $\eta$ differ by a factor of $\sim 3$ between the two analyses.
However, this arises because {\Fa} calculate a bolometric ratio, while our $\eta$ gives the ratio at a given energy. As the CRE spectrum is cut off by cooling, the overall energy in the $>1\GeV$ CREs (as accounted for by {\Fa}) underestimates the acceleration efficiency.

There are nevertheless some differences between the two studies.
The main one being that {\Fa} find a small (yet insignificant) positive signal from the FBs, whereas we find a negative (also insignificant) signal. This is due to our estimate relying on the extended distribution of the arrival directions, while {\Fa} counted the number of events whose center falls within the FBs.

Our analysis generalizes this neutrino--FB overlap by also examining limb-brightened emission from the shell, and utilizing the anticipated spectral hardness with respect to the background.
Finally, the edges used in the two analyses are somewhat different.

\acknowledgements
We thank N. K. Neilson, N. Whitehorn, E. Waxman, K. Murase, and K. Fang for helpful advice.
This research was supported by the ISF within the ISF-UGC joint research program (grant No. 504/14) and by the GIF (grant I-1362-303.7/2016), and received funding from the IAEC-UPBC joint research foundation (grant 257).

\bibliography{FermiBubbles}

\begin{thebibliography}{}
\expandafter\ifx\csname natexlab\endcsname\relax\def\natexlab#1{#1}\fi

\bibitem[{{Aartsen} {et~al.}(2014){Aartsen}, {Ackermann}, {Adams}, {Aguilar},
  {Ahlers}, {Ahrens}, {Altmann}, {Anderson}, {Arguelles}, {Arlen}, \&
  et~al.}]{IceCube37events}
{Aartsen}, M.~G., {Ackermann}, M., {Adams}, J., {et~al.} 2014, Physical Review
  Letters, 113, 101101

\bibitem[{{Aartsen} {et~al.}(2015){Aartsen}, {Abraham}, {Ackermann}, {Adams},
  {Aguilar}, {Ahlers}, {Ahrens}, {Altmann}, {Anderson}, {Archinger}, \&
  et~al.}]{IceCube15_SoftBackground}
{Aartsen}, M.~G., {Abraham}, K., {Ackermann}, M., {et~al.} 2015, \apj, 809, 98

\bibitem[{{Ackermann} {et~al.}(2014){Ackermann}, {Albert}, {Atwood}, {Baldini},
  {Ballet}, {Barbiellini}, {Bastieri}, {Bellazzini}, {Bissaldi}, {Blandford},
  {Bloom}, {Bottacini}, {Brandt}, {Bregeon}, {Bruel}, {Buehler}, {Buson},
  {Caliandro}, {Cameron}, {Caragiulo}, {Caraveo}, {Cavazzuti}, {Cecchi},
  {Charles}, {Chekhtman}, {Chiang}, {Chiaro}, {Ciprini}, {Claus},
  {Cohen-Tanugi}, {Conrad}, {Cutini}, {D'Ammando}, {de Angelis}, {de Palma},
  {Dermer}, {Digel}, {Di Venere}, {Silva}, {Drell}, {Favuzzi}, {Ferrara},
  {Focke}, {Franckowiak}, {Fukazawa}, {Funk}, {Fusco}, {Gargano}, {Gasparrini},
  {Germani}, {Giglietto}, {Giordano}, {Giroletti}, {Godfrey}, {Gomez-Vargas},
  {Grenier}, {Guiriec}, {Hadasch}, {Harding}, {Hays}, {Hewitt}, {Hou},
  {Jogler}, {J{\'o}hannesson}, {Johnson}, {Johnson}, {Kamae}, {Kataoka},
  {Kn{\"o}dlseder}, {Kocevski}, {Kuss}, {Larsson}, {Latronico}, {Longo},
  {Loparco}, {Lovellette}, {Lubrano}, {Malyshev}, {Manfreda}, {Massaro},
  {Mayer}, {Mazziotta}, {McEnery}, {Michelson}, {Mitthumsiri}, {Mizuno},
  {Monzani}, {Morselli}, {Moskalenko}, {Murgia}, {Nemmen}, {Nuss}, {Ohsugi},
  {Omodei}, {Orienti}, {Orlando}, {Ormes}, {Paneque}, {Panetta}, {Perkins},
  {Pesce-Rollins}, {Petrosian}, {Piron}, {Pivato}, {Rain{\`o}}, {Rando},
  {Razzano}, {Razzaque}, {Reimer}, {Reimer}, {S{\'a}nchez-Conde}, {Schaal},
  {Schulz}, {Sgr{\`o}}, {Siskind}, {Spandre}, {Spinelli}, {Stawarz}, {Strong},
  {Suson}, {Tahara}, {Takahashi}, {Thayer}, {Tibaldo}, {Tinivella}, {Torres},
  {Tosti}, {Troja}, {Uchiyama}, {Vianello}, {Werner}, {Winer}, {Wood}, {Wood},
  \& {Zaharijas}}]{FermiBubbles14}
{Ackermann}, M., {Albert}, A., {Atwood}, W.~B., {et~al.} 2014, \apj, 793, 64

\bibitem[{{Ahlers} \& {Murase}(2014)}]{AhlersMurase14}
{Ahlers}, M., \& {Murase}, K. 2014, \prd, 90, 023010

\bibitem[{{Bland-Hawthorn} \& {Cohen}(2003)}]{BlandHawthornAndCohen03}
{Bland-Hawthorn}, J., \& {Cohen}, M. 2003, \apj, 582, 246

\bibitem[{{Blattnig} {et~al.}(2000){Blattnig}, {Swaminathan}, {Kruger}, {Ngom},
  \& {Norbury}}]{BlattnigEtAl00}
{Blattnig}, S.~R., {Swaminathan}, S.~R., {Kruger}, A.~T., {Ngom}, M., \&
  {Norbury}, J.~W. 2000, \prd, 62, 094030

\bibitem[{{Carretti} {et~al.}(2013){Carretti}, {Crocker}, {Staveley-Smith},
  {Haverkorn}, {Purcell}, {Gaensler}, {Bernardi}, {Kesteven}, \&
  {Poppi}}]{CarrettiEtAl13}
{Carretti}, E., {Crocker}, R.~M., {Staveley-Smith}, L., {et~al.} 2013, \nat,
  493, 66

\bibitem[{{Cheng} {et~al.}(2011){Cheng}, {Chernyshov}, {Dogiel}, {Ko}, \&
  {Ip}}]{ChengEtAl11}
{Cheng}, K.-S., {Chernyshov}, D.~O., {Dogiel}, V.~A., {Ko}, C.-M., \& {Ip},
  W.-H. 2011, \apjl, 731, L17

\bibitem[{{Crocker} \& {Aharonian}(2011)}]{CrockerAharonian11}
{Crocker}, R.~M., \& {Aharonian}, F. 2011, Physical Review Letters, 106, 101102

\bibitem[{{Crocker} {et~al.}(2015){Crocker}, {Bicknell}, {Taylor}, \&
  {Carretti}}]{CrockerEtAl15}
{Crocker}, R.~M., {Bicknell}, G.~V., {Taylor}, A.~M., \& {Carretti}, E. 2015,
  \apj, 808, 107

\bibitem[{{Dermer}(1986)}]{DermerCrossSection}
{Dermer}, C.~D. 1986, \apj, 307, 47

\bibitem[{{Dobler} {et~al.}(2010){Dobler}, {Finkbeiner}, {Cholis}, {Slatyer},
  \& {Weiner}}]{DoblerEtAl10}
{Dobler}, G., {Finkbeiner}, D.~P., {Cholis}, I., {Slatyer}, T., \& {Weiner}, N.
  2010, \apj, 717, 825

\bibitem[{{Ellison} {et~al.}(2010){Ellison}, {Patnaude}, {Slane}, \&
  {Raymond}}]{SNRratio2010}
{Ellison}, D.~C., {Patnaude}, D.~J., {Slane}, P., \& {Raymond}, J. 2010, apj,
  712, 287

\bibitem[{{Fang} {et~al.}(2017){Fang}, {Su}, {Linden}, \& {Murase}}]{newlimit}
{Fang}, K., {Su}, M., {Linden}, T., \& {Murase}, K. 2017, ArXiv e-prints,
  arXiv:1704.03869

\bibitem[{{Finkbeiner}(2004)}]{Finkbeiner04}
{Finkbeiner}, D.~P. 2004, \apj, 614, 186

\bibitem[{{Fox} {et~al.}(2015){Fox}, {Bordoloi}, {Savage}, {Lockman},
  {Jenkins}, {Wakker}, {Bland-Hawthorn}, {Hernandez}, {Kim}, {Benjamin},
  {Bowen}, \& {Tumlinson}}]{FoxEtAl15}
{Fox}, A.~J., {Bordoloi}, R., {Savage}, B.~D., {et~al.} 2015, \apjl, 799, L7

\bibitem[{{Fujita} {et~al.}(2013){Fujita}, {Ohira}, \&
  {Yamazaki}}]{FujitaEtAl13}
{Fujita}, Y., {Ohira}, Y., \& {Yamazaki}, R. 2013, \apjl, 775, L20

\bibitem[{{Guo} \& {Mathews}(2012)}]{GuoMathews12}
{Guo}, F., \& {Mathews}, W.~G. 2012, \apj, 756, 181

\bibitem[{{Guo} {et~al.}(2012){Guo}, {Mathews}, {Dobler}, \& {Oh}}]{GuoEtAl12}
{Guo}, F., {Mathews}, W.~G., {Dobler}, G., \& {Oh}, S.~P. 2012, \apj, 756, 182

\bibitem[{{Hillas}(1985)}]{Hillas}
{Hillas}, A.~M. 1985, {The Origin of Ultra-High-Energy Cosmic Rays} (High
  Energy Astrophysics, Edited by Frederick Lamb. A Volume in the Annual Reviews
  Special Collections Program. Menlo Park, Calif. : Benjamin/Cummings Pub. Co.,
  c1985., p.277), 277

\bibitem[{{IceCube Collaboration}({2001})}]{IceCubeManual}
{IceCube Collaboration}. {2001}, { IceCube Preliminary Design Document}, Tech.
  Rep. {Revision 1.24}

\bibitem[{{IceCube Collaboration}(2006)}]{IceCubeIntro}
---. 2006, Astroparticle Physics, 26, 155

\bibitem[{{IceCube Collaboration}(2013)}]{IceCube28events}
---. 2013, Science, 342, 1242856

\bibitem[{{Kelner} {et~al.}(2006){Kelner}, {Aharonian}, \&
  {Bugayov}}]{rationumbers}
{Kelner}, S.~R., {Aharonian}, F.~A., \& {Bugayov}, V.~V. 2006, prd, 74, 034018

\bibitem[{{Keshet} \& {Gurwich}(2017{\natexlab{a}})}]{KeshetGurwich17a}
{Keshet}, U., \& {Gurwich}, I. 2017{\natexlab{a}}, \apj, 840, 7

\bibitem[{{Keshet} \& {Gurwich}(2017{\natexlab{b}})}]{KeshetGurwich17b}
---. 2017{\natexlab{b}}, ArXiv e-prints, arXiv:1704.05070

\bibitem[{{Kopper} {et~al.}(2015){Kopper}, {Giang}, \&
  {Kurahashi}}]{IceCube53EventsShort}
{Kopper}, C., {Giang}, W., \& {Kurahashi}, N. 2015, in International Cosmic Ray
  Conference, Vol.~34, 34th International Cosmic Ray Conference (ICRC2015), ed.
  A.~S. {Borisov}, V.~G. {Denisova}, Z.~M. {Guseva}, E.~A. {Kanevskaya}, M.~G.
  {Kogan}, A.~E. {Morozov}, V.~S. {Puchkov}, S.~E. {Pyatovsky}, G.~P.
  {Shoziyoev}, M.~D. {Smirnova}, A.~V. {Vargasov}, V.~I. {Galkin}, S.~I.
  {Nazarov}, \& R.~A. {Mukhamedshin}, 1081

\bibitem[{{Lacki}(2014)}]{Lacki14}
{Lacki}, B.~C. 2014, \mnras, 444, L39

\bibitem[{{Miller} \& {Bregman}(2016)}]{MillerBregman16}
{Miller}, M.~J., \& {Bregman}, J.~N. 2016, \apj, 829, 9

\bibitem[{{Morlino} \& {Caprioli}(2012)}]{MorlinoCaprioli12}
{Morlino}, G., \& {Caprioli}, D. 2012, aap, 538, A81

\bibitem[{{Mou} {et~al.}(2014){Mou}, {Yuan}, {Bu}, {Sun}, \& {Su}}]{MouEtal14}
{Mou}, G., {Yuan}, F., {Bu}, D., {Sun}, M., \& {Su}, M. 2014, \apj, 790, 109

\bibitem[{{Sarkar} {et~al.}(2015){Sarkar}, {Nath}, \& {Sharma}}]{SarkarEtAl15}
{Sarkar}, K.~C., {Nath}, B.~B., \& {Sharma}, P. 2015, \mnras, 453, 3827

\bibitem[{{Su} {et~al.}(2010){Su}, {Slatyer}, \& {Finkbeiner}}]{SuEtAl10}
{Su}, M., {Slatyer}, T.~R., \& {Finkbeiner}, D.~P. 2010, \apj, 724, 1044

\bibitem[{{Yang} {et~al.}(2013){Yang}, {Ruszkowski}, \& {Zweibel}}]{YangEtAl13}
{Yang}, H.-Y.~K., {Ruszkowski}, M., \& {Zweibel}, E. 2013, \mnras, 436, 2734

\bibitem[{{Zubovas} \& {Nayakshin}(2012)}]{ZubovasNayakshin12}
{Zubovas}, K., \& {Nayakshin}, S. 2012, \mnras, 424, 666

\end{thebibliography}

\end{document}